# A study into the impact of anti-extradition bill protests on Bangladeshi immigration into Hong Kong

Siddhartha Datta[1]

**Abstract:** Consequences from the 2019 anti-extradition protests in Hong Kong have been studied in many facets, but one topic of interest that has not been explored is the impact on the immigration of Bangladeshi immigrants into the city. This paper explores the value add of Bangladeshis to the Hong Kong, how the protests affected their mentality and consequently their immigration, and potentially longer-term detrimental effects on the city.

Looking through Hong Kong history, there is no shortage of anecdotes of the heroics of gurkhas, the contributions of police officers of Hindu or Sikh origin, or the noticeable differences made by Zoroastrian financiers [2]. But where are the Bangladeshis? It is a sombre irony that the first members from Bengal arrived in Hong Kong as low-class seamen the same way many Bangladeshi descendants illegally enter Hong Kong in boats at night [11, 12]. In the 20th century, including post-independence[2], there has been minimal references to Bangladeshi migration to Hong Kong. Under the premise of looking at this minor and possibly forgotten member of society, this essay intends to investigate the impact of the 2019 Hong Kong anti-extradition bill protests on the immigration of Bangladeshi nationals into Hong Kong. This essay is structured in three parts: (i) developing an understanding of the background of Bangladeshi immigrants in Hong Kong (HK), their incentive structures, and their contributions to HK society; (ii) uncovering the effect of the anti-extradition bill (AEB) protests on Bangladesh-HK (BG-HK) immigration and terminal effect on HK society; (iii) proposing policy recommendations to mitigate issues.

---

[1] sdatta@connect.ust.hk
[2] 1971-Present

# I. Bangladeshi immigrants in Hong Kong

## I.I. Context & Stakeholders in the immigration of a Bangladeshi.

The purpose of this section is to establish the context and narrow the stakeholder considerations when considering who had been impacted by AEB protests and how. For the immigration and value creation of a Bangladeshi individual, there are 2 key stakeholders directly involved in the process: the immigrant(s) (and dependents[3]), and the employer(s). The main differentiator in each of their roles, actions, and potential outcomes is first and foremost based on a single criteria: Was the immigration legal or illegal?

| Immigrants | <ul><li>Legal immigration (workers)</li><li>Illegal immigration (workers)</li></ul> |

Before we can establish the value add to society by Bangladeshi immigrants, we should try to understand the dynamics/relationship between HK society and Bangladeshi immigrants, and understand the incentive structure of the latter.

*Fig 1: Matrix of payoffs between HK society & Bangladeshi immigrants; 100 points arbitrarily allocated (55+22+18+5=100); colour represents the risk level posed to HK society (Dark is good for HK, light is harmful for HK)*

|  | **BG immigrants (legal)** | **BG immigrants (illegal)** |
|---|---|---|
| **HK society/situation (favorable to immigrants)** | (55) opportunistic value-add | (22) milk the city |
| **HK society/situation (unfavorable to immigrants)** | (5) Bangladesh retirement | (18) indifferent |

---

[3] Such as children, wives

The above matrix attempts to take a pseudo game theoretic approach in mapping out payoffs in 2 distinct scenarios for each player, the immigrant and the society. The immigrant is assumed to be able to choose between having legal status (by waiting the prerequisite time for working visa) and having illegal status (by taking a boat from China into HK while bypassing HK Immigration) [1]. Society encompasses attributes that may concern Bangladeshis/immigrants, such as employment, social status/stigma, social mobility, education for children, rule of law, etc, and can be deemed favourable or unfavourable towards immigrants. Whether society is considered to be "favourable" is relative to peer nations, comparison criteria could be arbitrarily based on, for example, intervening factors that propagate immigration entrepreneurship [9].

The following personal accounts[4] of Bangladeshi immigrants, obtained through discussions with the immigrants themselves (or in the case of *D*, through his peer) in brief interviews, can be used to validate or invalidate the above matrix.

*Fig 2: Personal accounts of 4 Bangladeshi immigrants of different statuses*

> *Immigrant A - Legal, Unfavourable*
>
> He grew up in a remote village in Bangladesh, and his parents and in-laws are based in the village. He came to HK in the early 2000s in his youth to earn money due to limited working opportunities in the village. He immigrated into HK legally with a valid visa, stayed with other single men in a shared apartment, and earned enough for basic survival. He sent most of his earnings back to his family in Bangladesh, taking advantage of the wide foreign exchange spread (1 HKD = 10 BDT). He was poorly educated and could not get work in mid-tier employment, mostly engaging in painting homes, plumbing, delivery, etc. Later on he got married, but could not afford to bring his wife and child to HK, so they stayed in Bangladesh while he sent money from HK to them for their survival. After a few years when he earned enough for a good

---

[4] The "personal approach" [10] is a ethnographic research technique used in methodology in which the author may introduce personal accounts of witnesses in extracting perspective-specific concepts to formulate an argument

standard living in Bangladesh after currency conversion from HKD to BDT, he took his funds and left HK to enjoy a better quality of life.

*Immigrant B - Illegal, Favourable*

He came to HK illegally, was not allowed to work legally, receives free supplies from the HK govt, but still works nonetheless for offices as cheap labour. He gives away supplies because he sometimes receives so much obsolescence supplies (e.g. excess soap). Because there is weak enforcement or check-ups, he can still work if he is careful. He has nothing to lose as he sends all his money back to Bangladesh for his family, and is not allowed to return back to Bangladesh.

*Immigrant C - Legal, Favourable*

He graduated from a city-level university in Bangladesh, has a Masters in Accounting, and is the sole breadwinner for his family living in HK. He has repeatedly attempted and failed the professional accounting license examinations (e.g. ACCA), as his qualifications are not recognized in HK to obtain accounting employment of standard pay. His educational qualifications have made it difficult for him to earn a wage as a professional competing with other local graduates. To cover expenses and keep his family in HK, he started a garments business in the mid-1990s. Although HK is not favourable in terms of employment, it has been quite favourable in terms of entrepreneurship opportunities. He was able to capitalize HK's close proximity to China and China's manufacturing capabilities, and earn an income comparable to Chinese peers who are more well off.

*Immigrant D - Illegal, Unfavourable[5]*

[This account was told through a peer] He claimed to have "swam" into HK (i.e. illegal immigration). He convinced his peers (both Chinese and Bangladeshis alike) to invest in a "lucrative" investment opportunity, and accumulated a large sum of money. One day he suddenly left HK along with all the money, and left many of his peers in financial burden. He had "married" a Korean girl in HK, but when he randomly left HK without giving all the debtors their money, he left the girl behind too. 5 years later, no one has received their money yet.

---

[5] The context under which an immigrant may be exerted may differ from other immigrants, depending on peer group, date of entry into HK, industry of employment, etc. Therefore we have both favourable and unfavourable conditions in these accounts.

For Immigrant A, he made the most out of the working opportunities and Forex differential between HK and Bangladesh, but HK does not offer him a good standard of living, so when the opportunity to leave arises (e.g. retired, or sufficient savings accumulated), he takes the opportunity to leave - this aligns with the "Bangladesh retirement" payoff.

For Immigrant B, HK is perceived to him as a temporary opportunity and he wishes to make the most out of it, but if things do not go well, he is willing to pay the consequences for any outcome.

For Immigrant C, he makes an active economic contribution to HK despite not being of comparable competitive standing to his peers. Bangladeshis trained within Bangladesh are at disadvantage in employment seeking, but his education or personal qualities still permit him to tap into other opportunities in HK that the city is willing to offer, and thus is able to find favour within the city. However, he lacks a sense of belonging to the city, and this may have other inadvertent social consequences.

For Immigrant D, he literally milked HK society and his own peers. He was engaged in a toxic environment in HK (no employment allowed, his illegal status, low respect/status amongst HKers as well as Bangladeshis), which may have led to his lack dignity and expectation of doing anything moral/ethical, and thus he was willing to engage in criminal behaviour. It appears he has no sense of belonging, not only to HK, but also to Bangladesh.

While these personal accounts are not representative of all the issues that arise, they highlight the circumstances (legal status vs favourability) that lead to certain social issues

(e.g. criminal behaviour, sense of belonging, entrepreneurship). These accounts help to validate the matrix as a tool for further decision-making or analysis in this essay. The matrix can subsequently be used to evaluate BD immigrant's value add to HK as well as how policy recommendations should be targeted.

| Employers | <ul><li>Employers of legal immigration</li><li>Employers of illegal immigration</li></ul> |
|---|---|

The perspective for an employer can be explained through prospect theory [7]. Employers are liable if they hire workers of illegal residency, whether intentionally or not. Illegal workers who are discovered to be working, both paid or unpaid, would be in violation of Immigration Ordinance section 38AA [4,5], and can be convicted of up to HK$50,000 fine with max 3 years imprisonment. For employers the fine can go up to HK$350,000 with also max 3 years imprisonment. Other than checking for HKID, employers are expected to confirm whether a job candidate has a recognisance form [4,5], a document that specifically prohibits the individual from working.

The payoff curve is clearly unbalanced between the potential reward an employer can derive from the worker versus the potential cost the employer may incur both financially legally and financially. As time passes, the cost side becomes more heavily weighted because the likelihood of discovery increases, thus in the long run the expected value of hiring an illegal worker is most likely negative. An additional constraint is the knowledge of whether a Bangladeshi worker had legally or illegally immigrated is ambiguous to the employer; documents can be forged and cases exist where employers have been held liable for employees with forged documentation [6]. Given this unknown uncertainty, the cost component becomes even greater than the reward component, thus eventually from a legal status perspective it appears to be in the rationally best interest of employers to avoid immigrant workers.

This is not the sole issue, but one perspective that employers are expected to hold with respect to immigration status. In the case that the skillset of the worker outperforms all other candidates, or the job requirements can be fulfilled by this individual several deviations better than other applicants, then the reward component may be distorted and may be close to balancing or even exceed the cost component.

Approximation of net value of hiring a Bangladeshi worker:

$$V(x,t) = p(legal) * R(x,t) - p(illegal) * C(x,t)$$

where function V(x) refers to the net value of hiring the worker; p(legal) refers to the probability the worker is permitted to work in HK legally and equal to 1-p(illegal); x refers to the specific individual in question and their ability to generate reward R(x) for the employer but also incur cost C(x), both components which may change over time based on the true status.

## I.II. How a Bangladeshi contributes to HK.

*Past Value-add: Historical recap*

Historically in the 1800s, there was both a plus side and a minus side to a Bengali presence in the South China Sea [12, 13]. Many were professionally considered as merchants *"lascars"*, seamen who traded items in unofficial and sometimes official capacity for the Honourable East India Company (HEIC). Others may remember that they contributed to the sales of opium in China, and were stereotyped as drunken sailors. When China rendered opium sales illegal, the lascars brokered between HEIC and China, accepting opium from the British and selling to China as a middleman. We can see similar parallels to how Bangladeshis thrive in modern Hong Kong society, both as enterprising individuals (as employers bridging different markets) but also as a downtrodden class of society meddling in illegal activities, as we will further elaborate.

Other than historical traits passing onto today's generation, it may also be worth noting the long-lasting maritime connection between Bengal and China (where today immigrants use China as an intermediary to enter HK illegally). In exercising its colonial philosophy, the British exerted its sea power through the merchants (or sometimes pirates); even though they had no allegiance to the Queen, they were executing trades and switching hands of silver between the Chinese and the British for the benefit and under "request" of the British, i.e. the British exerted indirect rule upon these sailors. Despite regional restrictions on opium consumption, the British also did not see it as any impediment, and simply navigated around the geopolitical issues of Sino-British opium trade, thus maintaining strong strategic understanding and management of geopolitics.

*Present Value-add: Census data*

Based on certain inferences we can make from the 2016 population by-census [8], we can converge into identifying one of the major avenues of value creation for HK society by Bangladeshi immigrants (both legal and illegal) is in manual labour and elementary occupations (e.g. domestic cleaning, food preparation aid, construction work, garbage collection, street cleaning).

Both legal and illegal immigrants are included in the census, and there are roughly 3700 Bangladeshis in Hong Kong. Most Bangladeshis in HK are Male and of working age in their 25-40s (29.2% out of all Bangladeshis) with labour force participation >80%. Most are employees (91%), though some have their own businesses (4%), majority taking part in elementary occupations (39%). Most immigrants arrived recently and did not exist from colonial times. Although no definitive statistics were reported on the breakdown of legal status, based on domestic living conditions, we can estimate that at least 63% of Bangladeshis in Hong Kong are legal immigrants. We attain this approximation based on the notion that it is extremely difficult for illegal immigrants to be married in HK (they may not bring their wives to HK, or will face challenges in permission to marry in HK, or cannot bring wives from Bangladesh after coming to HK, etc), and Bangladeshis who live with parents or have families with children (~63%) were able to bring or start a family in HK thus have a higher likelihood of being legal immigrants. We further corroborate this with the housing arrangement, where 26% of Bangladeshis live in non-relative / shared households (illegal Bangladeshi immigrants are known to live with other "bachelors"[6] in shared homes).

---

[6] A term the Bangladeshi community labels illegal immigrants

*Future Value-add: Comments analysis*

Census data reveals the contributions Bangladeshi immigrants perform in today's society, and with the education system in HK training the children of these immigrants, it should be expected that they would possess competencies that would enable them to perform non-elementary occupations or pursue professional careers. If Chinese candidates and Bangladeshi candidates possess similar level of skill and requirement fit for a role, it could be argued that bias against Bangladeshi may play a significant role. Thus we perform a review of comments made on 2 articles about 2 young locally-educated Bangladeshis who were children of immigrants, and test the degree of openness of HK netizens to the professional career pursuits elicited by them.

*Fig 3: Comment review analysis, to derive the Bangladeshi's perceived value-add ability in professional industries in Hong Kong by Hong Kong netizens; indexed as Cx*

| Believe Bangladeshis can add value | Believe they cannot |
|---|---|
| (1) Premise: Bangladeshi to become physician ||
| C1<br><br>edwintan2809@******<br>The reporter who wrote this piece should dig deeper. I think it's a ploy to keep the cushy jobs in locals'hands. No reason to exclude based on Cantonese.<br>2 years ago<br><br>REPLY  👍  👎 10 | C2<br><br>dolce.w@******<br>Are we supposed to feel sorry for him?<br>What is the official language of Hong Kong?<br>What language does the majority of Hong Kong people speak?<br>2 years ago<br><br>REPLY  👍  👎 4 |

| C3 | C4 |
|---|---|
|  **aahlohk@******<br><br>Hong Kong is a Chinese city . This rule the "Cantonese" rule excludes putonghua speaking students . Simply go and look at both CUHK and HKU intake , less than 1% mainland students. The paradox is all courses are taught in English .<br>In addition over 25% of these courses are attended by Canadian, American and Australian born Chinese, who surprisingly haven't paid HK taxes.<br>Both medical schools continue to increase their annual intake of solely Cantonese speakers, not reflecting the significant Filipino , Bangladeshi, Punjabi and expat communities of patients.<br>Hong Kong Asia's World City !!!!!! Ha<br><br>2 years ago<br><br>REPLY   👍   👎 7 |  **ippee28@******<br><br>Those ethnic minorities who chose HK as their adopted home should realize that HK is first and foremost a Chinese city with ethnic Chinese account for 94% (higher if you exclude the 300,000+ foreign domestic maids) of the population. Therefore, having Chinese language proficiency is a necessity if they don't want to be leftovers.<br>.<br>Now, just imagine a French-born-and-raised person who wants to be a doctor in France but his/her French is only at grade 2 level.<br><br>2 years ago |

**(2) Premise: Bangladeshi to become lawmaker**

| C5 | C6 |
|---|---|
|  **namwah**<br><br>I hope she will succeed. She would be much better than ex-colonial types who have lived here for decades but refusing to learn the languages of "natives".<br><br>last year<br><br>REPLY   👍   👎 0 |  **Pacific Century**<br><br>This girl is too naïve to think that Beijing will ever allow non-Chinese to have a say in HK politics and governance. Non-Chinese are always welcome to legally work, do business and make money in HK. But to actively participate in local politics is another story. |

| C7 | C8 |
|---|---|
|  **johnfra**<br><br>Go for it Fariha, your language skills would definitely be an advantage.<br>However, no need to be an ethnic minority lawmaker but become a full fledged Hong Kong lawmaker as you are a full HongKonger who has the advantage of being multi-lingual through your own hard-work.<br>Best of luck and good wishes.<br><br>last year<br><br>REPLY   👍   👎 0 |  **johndeed88@******<br><br>This Decilaslig is not too smart.<br>Don't care if she is Bangladeshi, I wish her the best of luck. What she should not do is just focus on the welfare of her ethnic people, she should stand for all the people in Hong Kong. Please don't do what that Senator has done in the US and wear a Syrian dress, that shows you where your allegiance really lie, and if you allegiance is still to your parents home country first, why the hell are you here?<br>Always said, if immigrants want to live here and be Hong Kongers, be Hong Kongers but don't bring your troubled cultures here. And before those liberals start attacking me, and demand we respect other cultures. Treating women as second class citizens in some countries and even rape seem more cultural in other societies, don't see you defending those.<br>Be a Hong Konger, drop your parent's country, don't bring host countries issues and demand we have to respect them here, fight for all Hong Konger alike, with that, I wish you the best of luck!!!!<br><br>last year<br><br>REPLY   👍   👎 0 |

Though it would not be fair to generalize the views of all HKers based on a few hand-selected netizen comments, they help illustrate certain phenomenon that may exist in HK society, in the pattern of Fear, Defensive Mechanism, and Offensive Mechanism [18, 19, 20].

Regarding the premises, article 1 [16] is about a Bangladeshi male of 18-24 age range who describes his difficulty in applying for medical schools in Hong Kong due to the language barrier and elaborating on the deficiencies of the education system for ethnic minorities. Article 2 [17] is about a Bangladeshi female of 18-24 age range as well who describes her ambition to become an ethnic minority lawmaker and speaks about her vision.

*Fear*

First, we can note a "threat to national identity" sentiment from some comments [C2, C7]. These comments, whether in support or against the individual's pursuit of the respective career, tend to return back to their view of what Hong Kong's identity is. In this small sample, we see repeated mention of the Chinese dialect/language, Chinese race, and what they believe to be the core values of HK. In the presence of adapting or modifying HK's Chinese-centric culture, these netizens resist and wish to reinforce and preserve the status quo.

*Defensive mechanism*

Second, we can observe a presence of "stigma localization". Stigma localization here intends to mean that local Chinese HKers project the cultural impression of people from a specific country upon the HK residents/citizens whose origin may be from that country.

This may result in stigma towards certain races; for example in C8 the netizen assumes people from a certain origin may possess certain cultural features (e.g. terrorism) that are harmful to the values of HK society, and assumes the individual places their origin above HK and will make their cultural features a primary agenda. Such stigma may appear as double standards into HK life, such as white *gweilo* are considered to be wealthy and treated politely while ethnic minorities are projected to be in poverty both in their country of origin as well as in HK. Comment C5 shows the converse, where the netizen holds a stigma against the colonizers and projects it to all white people who may enter politics in HK (i.e. they are sort of inferring that a person of colour will be more likely to learn the language of HK than a Caucasian).

*Offensive mechanism*

Third, we can note hints of "employment protectionism" [C1, C3, C4, C6]. There are implicit social structures and possibly explicit institutional structures that prevent members of foreign origin from taking certain roles and preserving them for members of local origin. For example, remarks are made about civil servants requiring certain requirements (e.g. lawmakers need to be "Chinese" as per Beijing's standards, or doctors must be able to communicate in Cantonese). For members of non-native origin, it is a major obstacle to overcome such requirements which may not necessarily be required to perform the job, and one could argue they are arbitrarily set to protect certain pivotal roles in HK society for "pure-bred" Chinese.

> The implications from these phenomena start with low social mobility for Bangladeshis, and worse, capped value creation ability of Bangladeshis for HK society (i.e. these structures inhibit HK to leverage immigrated and locally-trained Bangladeshis to create uncapped value). Immigration of Bangladeshis is not being stopped, so if immigrants are coming and accepting residency in Hong Kong, it makes game theoretic sense (referring to the matrix in Fig 1) to let them contribute by dismantling the social and institutional structures that may inhibit their contribution. Thus the direction of policy recommendation may follow along one of opening up opportunities for contribution, particularly areas that were traditionally closed off to ethnic minorities.

## II. Impact of AEB on BG-HK Immigration

In this section we will look into how the AEB events in 2019 affected BG-HK immigration, and in turn how it may have affected HK society both in the short run and potentially long run.

**II.I. Migration flow.**

Based on worldwide migration stock data [13, 14, 15], we can graphically map out flow networks between countries before and after AEB, with specific focus on Bangladesh and Hong Kong. We can map out where individuals of specific origin have moved from one location to another, out of their source/home origin to a target/destination location.

Migration stock refers to the number of people from a known (foreign) origin living in a specific location at a given time; migration/immigration flow would be the change in migration stock population between 2 years. First, we plotted the migration stock data at 5 year time intervals from 1990 to 2019, with the left graph representing the spaced-out (spring layout) network, while the right graph (circular layout) arranges all countries in a circle to ease readability. Each circle/node refers to a country, and edges/lines refer to immigration between two countries. The shorter the line / the closer two nodes are, the higher the edge weight, i.e. the higher the magnitude of migration.

In general it looks like HKSAR became a major immigration hub along with US and Germany after 2005, having the dense, close relationships with a wide variety of countries. Specifically what we find is that HK had a more central position in 2015 (high centrality) as an immigration hub than in 2019, with less immigration lines and relationships becoming more distant. It helps us infer that AEB weakened immigration ties (in terms of number of immigrants) between HK and BG.

Empirically, 444 BG immigrants were recorded to have immigrated into HK in 2015; 575 in 2019. However, the edges / relationship lines factor in / normalize for the magnitude of other flows, so for visualization purposes we refer to the network graphs.

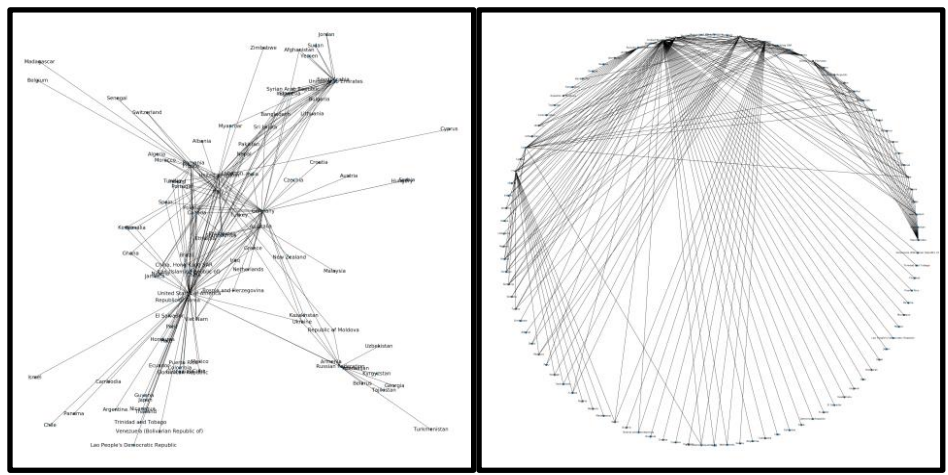

*Fig 4: Migrant Stock (local base of individuals); 2019*

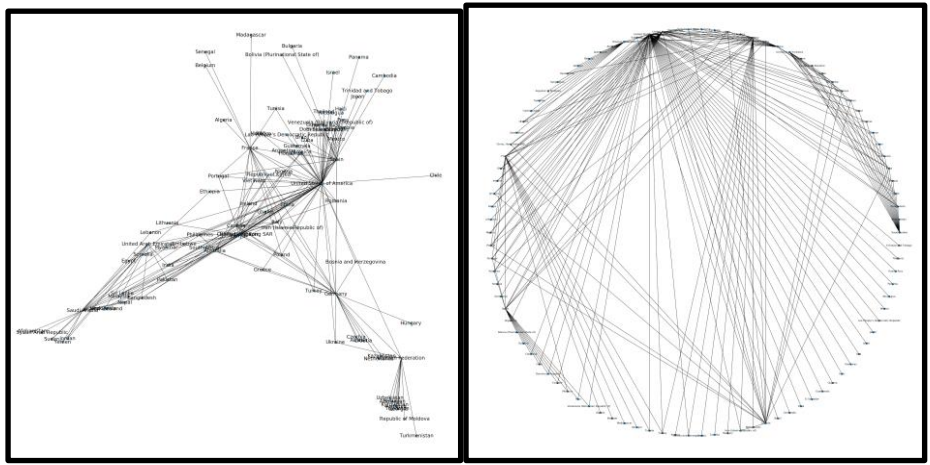

*Fig 5: Migrant Stock (local base of individuals); 2015*

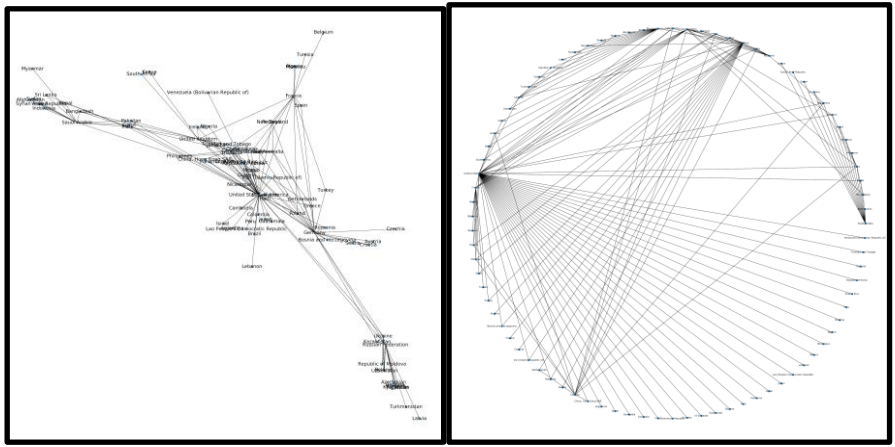

*Fig 6: Migrant Stock (local base of individuals); 2000*

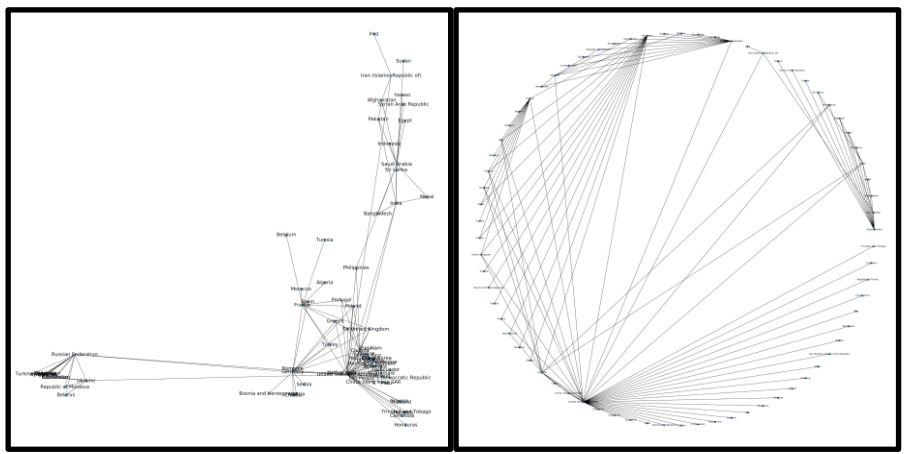

*Fig 7: Migrant Stock (local base of individuals); 1990*

Next, we wished to identify the different immigration source countries sending immigrants into HK; here, we are measuring immigrant flow directly (not migration stock). HK is marked in red, and we note that although Bangladesh is one of the top exporter of immigrants to HK, for comparison of scale we find that Philippines, Indonesia and Thailand send magnitudes more immigrants (which makes sense given these are the typical nationalities of HK's domestic helper base).

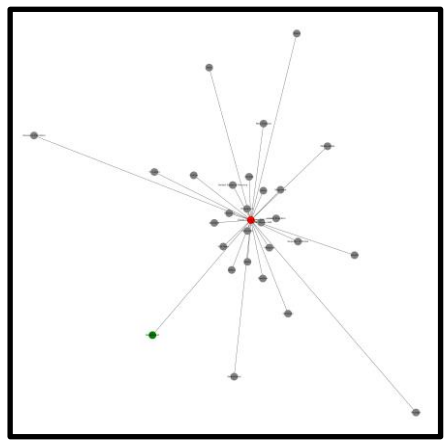

*Fig 8: All immigration sources into HK target; 2019-2015*

We also wished to see where the BG immigrants are going, so we plotted the destination countries for Bangladeshis (relative to 2015), and what we find is that HK (red) is further back from BG (green) than other countries, and BG immigrants have placed relative preference to countries such as US and Canada. At least in terms of neighbouring countries, Singapore or China did not enjoy a strong shift in immigration away from HK, so HK still holds a grip as Asia's optimal working destination for Bangladeshis. As a control, we also did the same to HK to see which countries HK people immigrate into out of HK, and we see China, South Africa and the US in the close destinations.

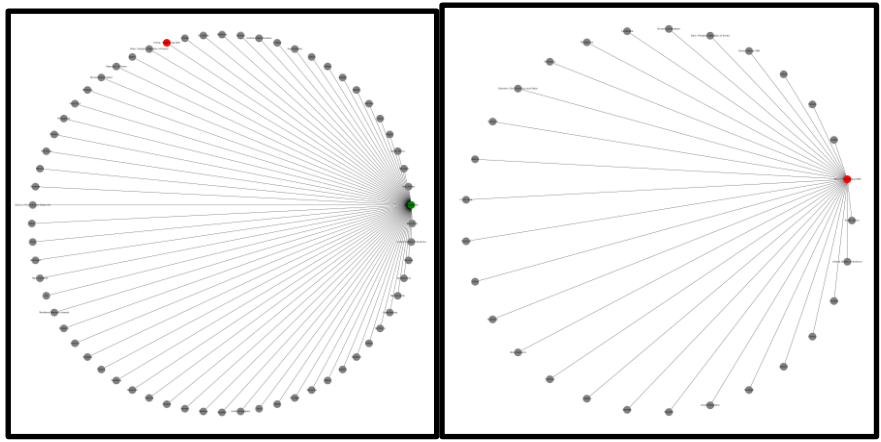

*Fig 9: Immigration out of Bangladesh as source node (left), Immigration out of Hong Kong as source node (control, right); 2019-2015*

**II.II. Implications on Hong Kong.**

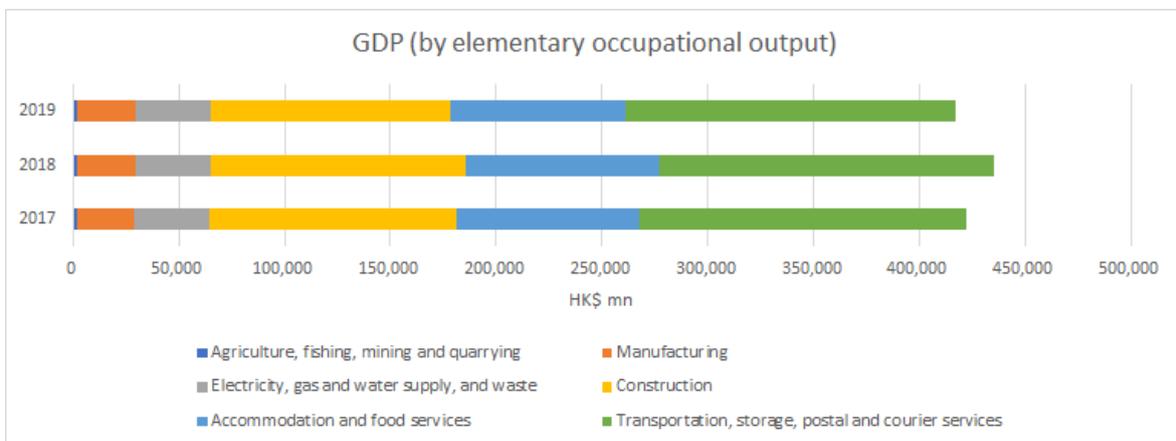

*Fig 10: GDP output per elementary occupational sector (HK$ mn) [21, 22]*

As established in the previous section, the majority of Bangladeshi's contribution to HK at present is in elementary occupations. Filtering GDP components for such occupations, we can observe a decline from 2018 to 2019. Specifically, construction and accommodation output faced the severest fall amongst the elementary occupational outputs, with -5.99% and -9.23% change between 2018 to 2019.

One could argue that this decline may have been caused by the general decline in economic performance in HK, for example attributable to hotels without occupancy or construction contracts being cancelled, etc. This may be true, so to justify that this worsened economic performance may be attributable to negative immigration, we will need to check whether there was labour shortage in these sectors.

Reports from the Construction Industry Council (CIC) [21, 22] verify that indeed there is and will be continued shortage in the range of 5,000-10,000 workers in the construction industry till 2022. This may support the notion that decreased general migrant flow, and possibly Bangladeshi migrant flow, might have contributed to this labour shortage that incurred irreversible systemic damage in the short run and requires an influx of immigrants to correct.

The aforementioned data on economic impact is presented to reflect the economic performance caused by AEB, but the direct causation between AEB and reduced immigration of Bangladeshis and labour shortage cannot be proven definitively at this stage. In fact, it is possible that elementary occupational GDP output is a lagging indicator to immigrant migrational flow, thus the true effect of AEB on GDP may not be fully realized in 2019 but in later years when labour shortage issues become more severe.

This may be the case given that CIC reported that amongst the 483,000 construction workers, 43.8% are aged and >50 while only 14.7% are aged below 30 and well-fit for construction work. This already is starting to indicate an incoming systemic issue in certain elementary occupations.

Furthermore, referring back to the HK society vs Bangladeshi matrix in Fig 1, one could argue that AEB protests have moved the optimal outcomes based on HK's situation from "favourable to immigrants" to "unfavourable to immigrants" (resulting in an eventual departure from HK); this societal position may still yield relatively similar level of outcome for illegal immigrants (indifferent), so there is a risk that HK will miss out on better talent of legal immigrants.

Reflecting on the migration flow data, although the change in flow rate ($\Delta f\%$) of Bangladeshi immigrants into Hong Kong is -17%, the actual flow rate ($f\%$) in Bangladesh immigrant base in Hong Kong in 2019 is still positive (+12%), i.e. Bangladesh immigrants are still coming to HK, just at a lower rate than before. Other than easing economic concerns slightly, the deeper implication from this is that Bangladeshis immigrating into Hong Kong have accepted HK's political situation and uncertainty, and are willing to show resilience, i.e. create incremental societal value, perhaps to the extent that Bangladeshis are integrating with HK society from day 1.

> To recap: AEB certainly has had an effect on BG-HK immigration, and though the effect onto HK is empirically evident to a medium extent, the possibility for the consequences of AEB and continued political instability to cause harm to HK society increases. Elementary economic contributions, labour shocks, potential overflow of non-value-adding immigrants, and loss of potential legal immigration talent to other Western countries.

# III. Policy recommendations

As shown in the matrix in Fig 1, it is most favourable for HK to maximize the value derived from those Bangladeshi immigrants who wish to contribute to HK opportunistically. This means we need to provide them opportunities to create value, and can do so with the below two policy recommendations.

**Policy 1: "Diversity Inclusion Programme".**

Unlike the US's Diversity Visas lottery scheme [23], HK's problem is not getting immigrants into the city, the problem is integrating the immigrants into society. There is a gap: (1) if they are coming to HK society anyways and consuming tax resources (i.e. they need to be trained to do work, they are not ready as-is), and (2) HK people are not multicultural (although we have EMs in HK, Chinese segregate themselves from EMs). The solution is thus the govt can subsidize companies to hire illegal as well as legal employees (and fast-track their citizenship status by 4 years), e.g. tax benefits or pay part of salary, etc. One might note that illegal immigrants under favourable conditions might push them to abuse the system, but if we grant them citizenship for their contributions and improved growth, it builds a sense of belonging to the city and move their value add to one of opportunistic value add (after they change from illegal to legal status).

A similar solution in the education field is implemented in Hong Kong local secondary schools, where the government subsidizes schools who take in a certain number of EM students [24]. The limitation of the education-based recommendation was that the subsidy can be abused to obtain funds but allocated however the school sees fit but not actually invested in ethnic minorities. Similarly, we might see firms accept funding, but rather than train the EM workers, they may just give them low rank roles and never promote them, and allocate the funds for other purposes. This means the government, if they were to implement the policy, would need to enforce it in the first few years with a defined set of components that companies need to use the funds for (e.g. employee training, Chinese-EM integration events, technical workshops).

**Policy 2: "Homemaker Educational Retraining Scheme".**

Most Bangladeshis or ethnic minorities who legally immigrate into HK come with their wives. Usually the males will be the breadwinners and start working, but the females are usually young and are housewives, and have time to invest in other self-improvement activities. The government or even NGOs can provide young EM housewives with educational subsidies / sponsorships / scholarships to study in universities or community colleges to improve their educational attainment (for example, most Bangladeshi women who came to HK in the 1990s with their husbands were aged around 19-21). This would help them also become earners for the family, pushes the agenda for gender equality values, and it also means housewives who raise children possess a stronger educational background and can act as both educational guide as well as role model to the child (e.g. it is uncommon for EM children to be parented by degree-holders).

In terms of overall impact, both policies are pursuing the strategy of activating legal immigrants and maximizing opportunities for them to leverage. As social features (e.g. social stigma) take time to phase out, instead institutions can commence with breaking down institutional structures biased against Bangladeshis and ethnic minorities, and the policies may also lead to changing the mindset of society towards EMs from a top-down

> approach. Illegal immigrants become legal ones, uneducated immigrants become educated ones, labour-inactive immigrants become labour-active; these are some of the changes in states that the above 2 policies will be able to execute, not only resolving labour shortage issues in HK's general workforce (not just elementary occupations), but also granting social mobility to  explore other career options.

**Key Conclusions & Message.** All in all, we talked about how each kind of Bangladeshi immigrant can create what type of value for HK, how AEB weakened their value creation ability in elementary occupations, and how the government could unlock opportunities to push Bangladeshis beyond just elementary work and create even more value in areas that may be better suited to them in the long run.

Bangladeshis in Hong Kong can be more than what they currently seem, they just need to be given the chance to prove so.

**Citations [References also included within essay]:**

[1] "Bangladeshi Man Falls to Death from Cliff after Entering City Illegally." 2017. South China Morning Post. https://www.scmp.com/news/hong-kong/law-crime/article/2095495/bangladeshi-man-falls-death-cliff-shortly-after-entering (June 17, 2020).

[2] "At Home with the Harilelas: A Look inside Family's Famous Mansion." 2019. South China Morning Post. https://www.scmp.com/lifestyle/article/2187314/inside-hong-kong-mansion-harilelas-one-citys-wealthiest-families (June 17, 2020).

[3] "Bengali Hindu Surnames." 2020. Wikipedia. https://en.wikipedia.org/w/index.php?title=Category:Bengali_Hindu_surnames&oldid=942413923 (June 17, 2020).[7]

[4] "Two Bangladeshi Illegal Workers Jailed | Immigration Department." https://www.immd.gov.hk/eng/press/press-releases/20160610.html (June 17, 2020).

[5] "Three Guns and HK$2.4 Million in Drugs Seized during Arrest of Man." 2020. South China Morning Post. https://www.scmp.com/news/hong-kong/law-and-crime/article/3064801/police-tracking-gang-ringleader-find-three-guns-and (June 17, 2020).

[6] "Twenty-Nine Persons Arrested during Anti-Illegal Worker Operations (with Photo) | Immigration Department." https://www.immd.gov.hk/eng/press/press-releases/20191025b.html (June 17, 2020).

[7] Kahneman, Daniel and Tversky, Amos. 1979. "Prospect Theory: An Analysis of Decision under Risk" *Econometrica* 47 (2): 263–291.

---

[7] Bengali names, used to track down history of pre-independence Bengal traces in Hong Kong

[8] Census and Statistics Department HKSAR. 2016. "Thematic Report: Ethnic Minorities" *Population By-Census 2016*[8]

[9] Nyga, Honorata and Eliza, Przezdziecka. 2017. "Modelling Energy Security and International Competitiveness: The Export Perspective". *Entrepreneurial Business and Economics Review* 5. 71. 10.15678/EBER.2017.050204.

[10] Honigmann, John J., Mina Davis Caulfield, Simeon W. Chilungu, Raymond Eches, Paul Wald, Anna-Britta Hellbom, Charles Keil, Hilda Kuper, L. L. Langness, Jacques Maquet, Dennison Nash, Pertti J. Pelto, Gretel H. Pelto, Gopala Sarana, Charles L. Siegel, Elvi Whittaker, and Rolf Wirsing. 1976. "The Personal Approach in Cultural Anthropological Research" *Current Anthropology* 17, no. 2: 243-61.

[11] "Who Were the Lascars – and Where Have They Gone?" 2020. South China Morning Post. https://www.scmp.com/magazines/post-magazine/short-reads/article/3044306/who-were-lascars-where-did-they-come-and-where (June 17, 2020).

[12] "The Start of Foreign Mud." https://gwulo.com/node/6165 (June 17, 2020).

Black, Parbury, and Allen. 1842. *The Asiatic Journal and Monthly Register for British India and Its Dependencies* Volume 39

[13] United Nations, Department of Economic and Social Affairs, Population Division. 2019. "International Migrant Stock 2019". *United Nations database* POP/DB/MIG/Stock/Rev.2019

[14] Datta, Siddhartha. 2020. dattasiddhartha/gflo. https://github.com/dattasiddhartha/gflo (June 17, 2020).

[15] Datta, Siddhartha. 2020. Immigration flow graphing. https://github.com/dattasiddhartha/gflo/tree/master/examples/immigration-flow (June 17, 2020).[9]

---

[8] The census accounts for asylum seekers / known illegal immigrants. They factor in mobile and non-permanent residents based on the "resident population" approach.

[9] Source code for the network analysis of immigration flow

[16] "Teen's Dreams Left in Tatters Because He Can't Speak Chinese." 2018. South China Morning Post. https://www.scmp.com/news/hong-kong/education/article/2154479/hong-kong-teens-dreams-being-doctor-dashed-education (June 17, 2020).

[17] "She's 20, Speaks 6 Languages – and Wants to Represent You." 2019. South China Morning Post. https://www.scmp.com/news/hong-kong/society/article/2180667/20-year-old-student-bangladeshi-origin-hopes-become-hong (June 17, 2020).

[18] Givens, Terri E. 2015. "The Racialization of Security: Ethnic Minorities in Europe." *International Relations and Comparative Politics*.

International Organization for Migration. 2001. "International Migration, Racism, Discrimination and Xenophobia."

[19] Constant, A. 2014. "Do migrants take the jobs of native workers?." *IZA World of Labor* 2014: 10 doi: 10.15185/izawol.10

[20] Lewis, Ethan. 2017. "How immigration affects workers: two wrong models and a right one." *The Cato Journal*, vol. 37, no. 3, 2017, p. 461-472.

[21] "Panel on Development LC Paper No. CB(1)1086/18-19(04)." https://www.legco.gov.hk/yr18-19/english/panels/dev/papers/dev20190528cb1-1086-4-e.pdf (June 17, 2020).

[22] "Construction Expenditure Forecast." http://www.cic.hk/common/Fore/disclaimer.aspx?lang=en-US&year=2018_19v2 (June 17, 2020).

[23] "Diversity Visa Program - Entry." https://travel.state.gov/content/travel/en/us-visas/immigrate/diversity-visa-program-entry.html (June 17, 2020).

[24] "More support for ethnic minorities." Hong Kong's Information Services Department. http://www.news.gov.hk/eng/2019/10/20191027/20191027_093632_116.html (June 17, 2020).